\documentclass[aps,prl,reprint,groupedaddress,longbibliography]{revtex4-1}
\usepackage{graphicx}

\begin{document}

\title{Metrics for two electron random potential systems}

\author{A. H. Skelt, R. W. Godby and I. D'Amico}
\affiliation{Department of Physics, University of York, York YO10 5DD, United Kingdom}

\date{\today}

\begin{abstract}
Metrics have been used to investigate the relationship between wavefunction distances and density distances for families of specific systems.  We extend this research to look at random potentials for time-dependent single electron systems, and for ground-state two electron systems.  We find that Fourier series are a good basis for generating random potentials.  These random potentials also yield quasi-linear relationships between the distances of ground-state densities and wavefunctions, providing a framework in which Density Functional Theory can be explored.
\end{abstract}

\maketitle

\section{Introduction}
\label{sec:intro}
Calculating many-body physics exactly is a great challenge due to the exponentially increasing numerical cost of storing the wavefunction with increasing particle number.  To overcome this problem, Density Functional Theory (DFT) \cite{Parr1994,Burke2012} uses the density as a basic variable to describe the system's ground-state properties.  At the heart of DFT lies the Hohenberg-Kohn theorem where the ground-state wavefunction and density are demonstrated to have a one-to-one mapping \cite{Hohenberg}.  Although this is a key concept for DFT, the relationship is not fully understood.  Here we follow on from previous research \cite{DAmico,S&D,S&D2,Skelt2017}, using these metrics to further understanding of this one-to-one mapping of density and wavefunction.

Metrics are used as a measure of distance, and in particular ``natural'' metrics have previously been used to study density distances and wavefunction distances.  The relationship between these two distances for families of closely related systems in their ground state was shown to be quasi-linear; linear for most distances, but deviating from this linear relationship close to the maximum distances to ensure these distances are reached together in the wavefunction and density \cite{DAmico,S&D}.  Recent work has extended this to show that this relationship extends beyond related systems to families of unrelated single-electron systems, and has been expanded further to look at time-dependent systems using the metrics.  This has been used to characterise quantum dynamics and adiabaticity in single-electron quantum systems \cite{Skelt2017}.

Here we show how to generate the random systems used in Ref.~\cite{Skelt2017}, ensuring the electrons remain confined whilst still providing a range of systems with rich physics to explore.  We then apply metrics to time dependent single electron systems, looking at the mapping between wavefunctions and densities.  As the number of electrons increases, the relationship between the wavefunction and the density gets increasingly more complicated, and so it is the natural progression to look at larger systems using these metrics.  Therefore we move on to applying these techniques to two-electron ground-state systems. We explore the wavefunction-density mapping for DFT using the metrics, and compare interacting and non-interacting results, which can be useful when considering Kohn-Sham systems in DFT \cite{Kohn1965}.

\section{Random potentials for quantum systems}
\label{sec:rand}
Much of the work looking at solving quantum systems exactly uses very well known systems, such as the Hubbard model or Hooke's atom \cite{DAmico,S&D,Herrera2017}.  However, the iDEA code has the ability to numerically exactly solve any 1, 2 or 3 electron system in one dimension for the density and the wavefunction \cite{Hodgson}.  We therefore look at methods of randomly generating a wide range of systems, ``random potentials'', to be solved exactly by iDEA.

By generating a wide range of systems, we can explore systems with diverse physics. There are not the constraints of just varying one or two parameters of the Hamiltonian, such as the frequency of the Hooke's atom.

However, we still require some constraints.  The potential must be confining overall, and the shape must be smooth so the system is still physical.  Using these requirements, we look at using polynomials and Fourier series with randomly generated coefficients to produce the random potentials.

\subsection{Polynomials}
\label{sec:poly}
To generate the random potentials using polynomials, we used polynomial series up to even powers, to ensure the confinement of the electrons.  Each term in the series was then assigned a random coefficient from a uniform distribution.  However it was seen that these were insufficient for producing varied secondary wells in the potential, `microwells', and the potential was predominantly flat.  This was exacerbated when an $x^{10}$ confining potential was applied to affirm the electron confinement.

\subsection{Fourier series}
\label{sec:fourier}
To improve on the polynomial series for obtaining random potentials, we turn to a Fourier series with random coefficients.  Since the electrons still need to be confined, an overall potential of $x^{10}$ is applied, but scaled by $1/10^{11}$ to ensure the oscillations from the Fourier series are still dominant.  An $x^2$ confining potential was also investigated but it was found to be too dominating at the edges of the system.

The form of the Fourier series is:
\begin{equation}
\label{eq:rand_pot}
V_{ext} = \frac{x^{10}}{10^{11}} + \Lambda \sum_{n=1}^3 \left( a_n \cos \frac{n \pi x}{L} + b_n \sin \frac{n \pi x}{L} \right),
\end{equation}
where $\Lambda$ is the confining strength of the potential microwells, $L$ is half the system size (in atomic units\footnote{We denote atomic units using a.u., where $\hbar = m = e = 4 \pi \varepsilon_0 = 1 $.}) and $a_n$ and $b_n$ are random numbers from a uniform distribution ranging from $-L/3$ to $L/3$.

\begin{figure}
\includegraphics[width=0.48\textwidth]{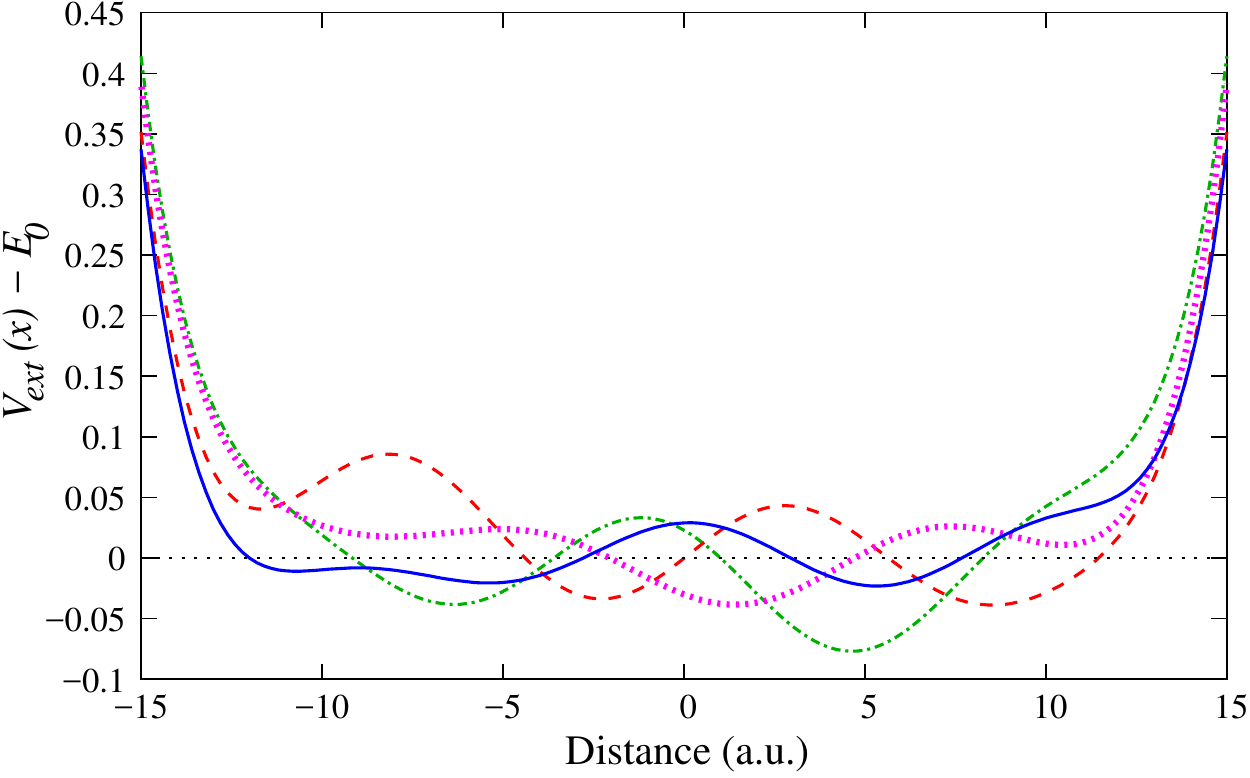}
  \caption{A range of potentials generated using Eq.~\ref{eq:rand_pot}, shifted so the ground-state energy lies at 0 (shown by the horizontal dotted line).}
\label{fig:pots}
\end{figure}
As we can see from Fig.~\ref{fig:pots}, Eq.~\ref{eq:rand_pot} yields a wide variety of confining potentials.  Note that these potentials all have three microwells due to Eq.~\ref{eq:rand_pot} being a Fourier series up to 3 terms of $\cos(x)$ and $\sin(x)$.  For more wells, the series just needs to be extended to the number of microwells desired. To increase (decrease) the tunneling between wells, $\Lambda$ can be decreased (increased).  The system size can also be increased to accommodate more microwells if necessary.  Eq.~\ref{eq:rand_pot} is a versatile method for generating a wide range of physical quantum systems.

We see that this formulation gives a wide range of potentials leading to diverse static and dynamic behaviours.  This way of generating systems, therefore, enables broader studies of quantum phenomena.  This is enhanced by the iDEA code's ability to exactly solve any potential for one, two or three electron systems.

\section{Metrics}
\label{sec:mets}
Metrics are a useful way to measure the distance between quantities.  In Ref.~\cite{DAmico}, ``natural'' metrics for calculating the wavefunction distance and density distance were developed from conservation laws.  While other metrics could in principle be used, this protocol for deriving metrics ensures that they are at the same time not arbitrary and physically sound.  These ``natural'' metrics are:

\begin{equation}
  \label{eq:wf_met}
  D_{\psi}\left(\psi_1, \psi_2 \right) = \left[ 2N - 2 \left| \int \psi^*_1 \psi_2 \, dr_1 \ldots dr_N \right| \right]^{\frac{1}{2}} \, ;
\end{equation}
\begin{equation}
\label{eq:den_met}
  D_{n}\left( n_1, n_2 \right) = \int |n_1(\textbf{r}) - n_2(\textbf{r})|d^3\textbf{r} \, .
\end{equation}

It has been studied how these metrics can show the relationship between ground state densities and wavefunctions for families of systems where one parameter is varied \cite{DAmico,S&D}.  These results showed that there was a quasi-linear relationship for these related systems, the gradient of which was dependent on the number of electrons, $N$ (up to $N=8$) \cite{DAmico}.  More recent research has extended this to show that this relationship holds beyond related families of systems, to families of unrelated, random potentials of one electron.  This is then used to determine adiabaticity in quantum systems \cite{Skelt2017}.

We will extend the use of metrics to look at this relationship for single electron random systems under very fast dynamics.  This opens up questions about the ergodicity of wavefunctions and densities in metric space.

We will also see here how this relationship also holds for two-electron random systems in their ground-state, providing a good basis for the use of metrics to compare any two systems of $N$-electrons in their ground-state.  This can then be used for investigating the one-to-one mapping of the density and wavefunction, at the heart of DFT, or for characterising quantum dynamics and adiabaticity (as has already been shown for single electron systems), among other applications.

\section{Time-dependent single electron random-potential systems}
\label{sec:1e}
\begin{figure}
\includegraphics[width=0.48\textwidth]{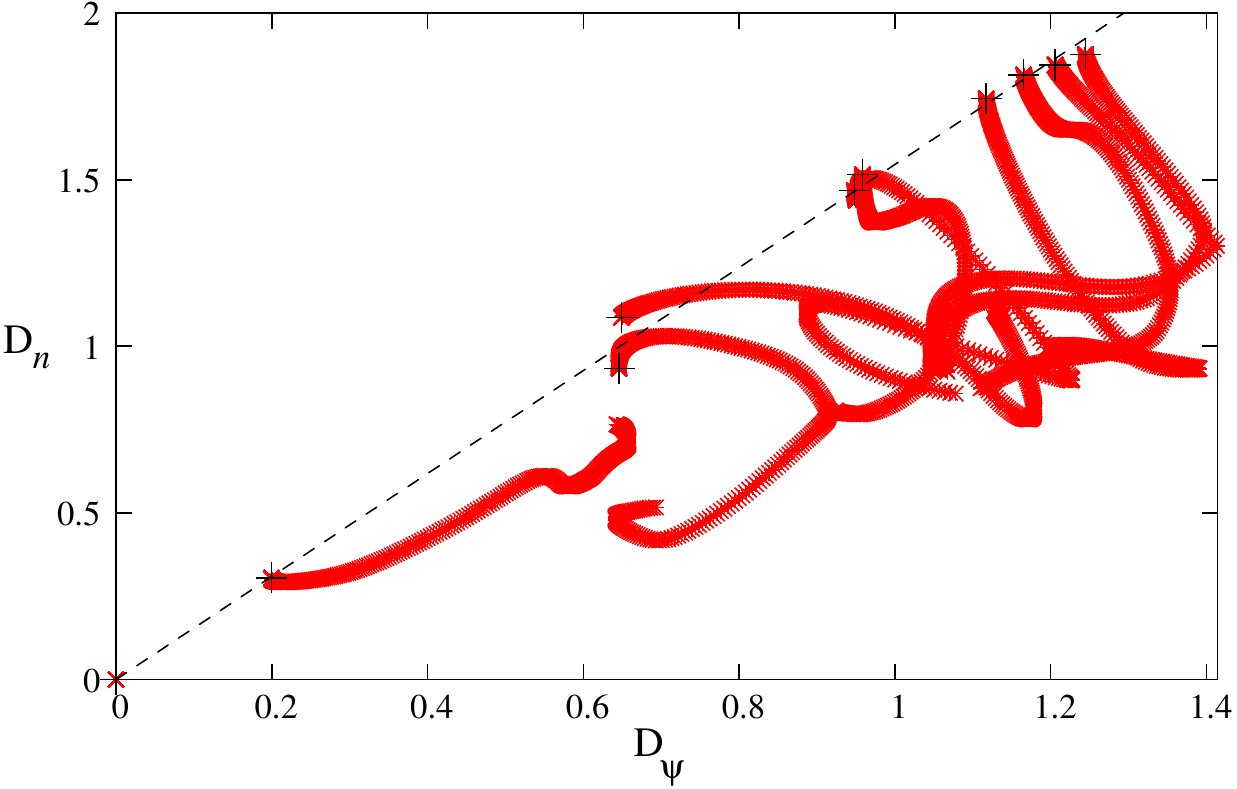}
  \caption{Density distance ($D(n_1(x,t),n_2(x,t))$) against wavefunction distance ($D(\psi_1(x,t),\psi_2(x,t))$) for a family of random systems experiencing an electric field of strength 0.01 a.u. at $t=0$. The field induces a strongly out-of-equilibrium dynamics. The `ground-state line', along which the systems would align if in the ground-state or if adiabatic, is shown by the black dashed line ($y=1.55x$). The distances between the ground-states at $t=0$ are marked by the black crosses, and indeed can be seen to lay close to the ground-state line. Once the perturbation is applied, the distances, seen by the red trails, no longer remain on the ground-state line.}
\label{fig:1e_td}
\end{figure}

In this section, we apply the metrics to a range of random potentials (see Table~\ref{tab:sys_param} in Appendix~\ref{sec:append} for the parameters used in Eq.~\ref{eq:rand_pot} to generate these random potentials), focusing on the time-dependent single-electron systems.  It is noted that for ground-state single-electron systems, the linear relationship is seen with a gradient of $\sim$1.5 \cite{Skelt2017}, as reported by the black dashed line in Fig.~\ref{fig:1e_td}

A family of 10 random single electron systems is perturbed by a constant, uniform, electric field of strength 0.01 a.u. at $t=0$. This relatively strong electric field creates a current flow that is highly position- and time-dependent. These systems' dynamics were compared to a reference system within the family, at each time step.  Fig.~\ref{fig:1e_td} shows how the systems evolve with respect to each other, in terms of the density distance and wavefunction distance. 

Under this highly out-of-equilibrium regime, the ratio between the density distance and wavefunction distance (between any two systems) does not remain linear, but in fact begins to explore the region below the ground-state line (Fig.~\ref{fig:1e_td}).  This suggests a non-ergodic nature of quantum dynamics in metric spaces. Also from this exploration of the lower triangle, it can be seen that the wavefunction distance is, on average, affected more than the density distance under out-of-equilibrium dynamics.

There is no guarantee that two similar densities, with a small distance between them, correspond to two similar wavefunctions, and the exploration of this lower triangle corresponds to these situations.  However the reverse of two similar wavefunctions corresponding to two very different densities, which would lead to the upper triangle being explored, is not observed.\footnote{The Runge-Gross theorem does not prohibit two systems, propagated by (possibly different) time-dependent potentials from different initial many-body states from having the same density at some later instant. Such a case would provide a trajectory that touches the very bottom of the lower triangle in Fig.~\ref{fig:1e_td}: $D_n=0$ while $D_{\psi} \neq 0$. This provides further support for time-evolution in the metric space of Fig.~\ref{fig:1e_td}. being predominantly concentrated away from the upper triangle.}  This provides intricate details of the mapping between densities and wavefunctions.

\section{Ground-state two electron random-potential systems}
\label{sec:2e}
It was seen that the ground-state quasi-linear relationship between wavefunction distances and density distances was not just for families of related systems, but also indeed holds for families of single-electron random systems.  We now investigate whether this is the case for systems of 2 electrons with random potentials (see Table~\ref{tab:sys_param} in Appendix~\ref{sec:append} for the parameters used in Eq.~\ref{eq:rand_pot} to generate these random potentials), for both interacting and non-interacting systems.

It must be noted that Eqs.~\ref{eq:wf_met} and \ref{eq:den_met} are normalised to the number of electrons, $\sqrt{N}$ and $N$ respectively, but Figs.~\ref{fig:2e_int} and \ref{fig:2e_non_int} normalise the metrics to 1.  This is for ease of comparison with the quasi-linear relationships found in Ref.~\cite{DAmico}.

\subsection{Interacting systems}
\label{sec:int}
\begin{figure}
\includegraphics[width=0.48\textwidth]{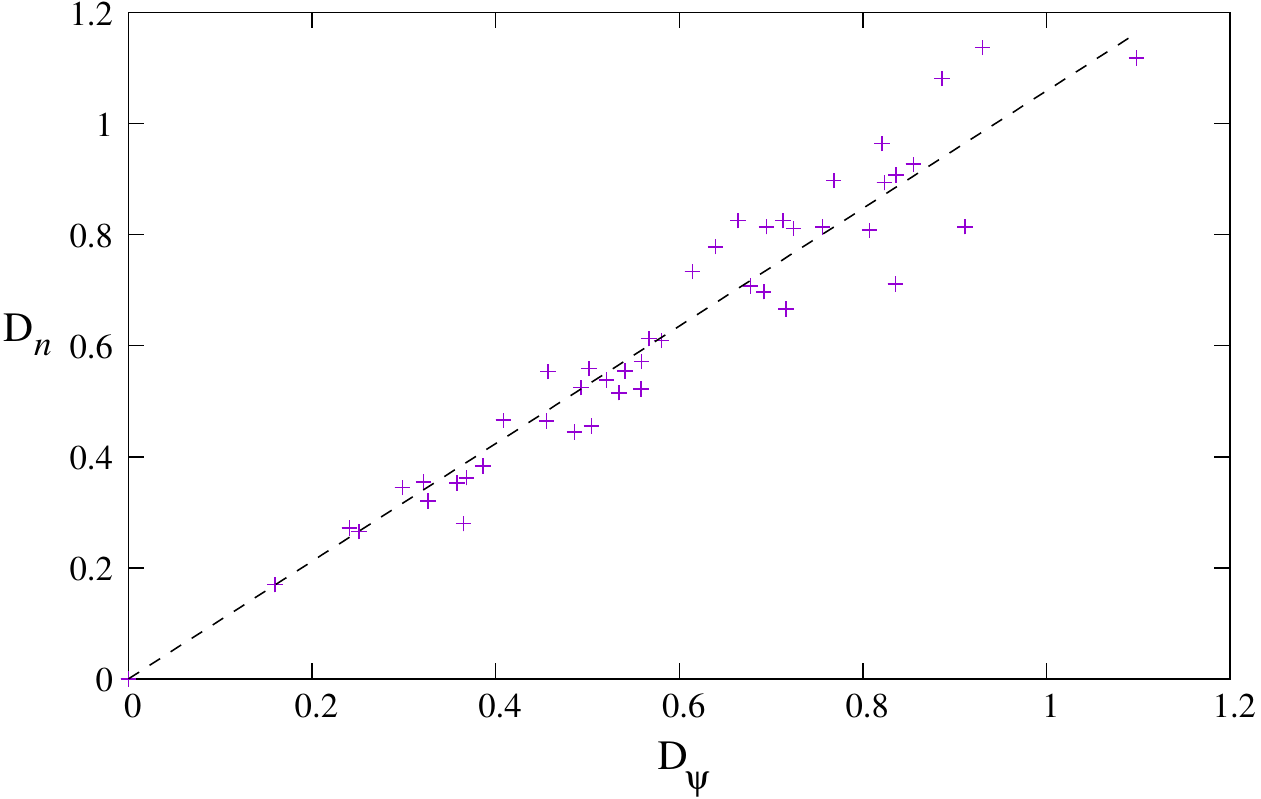}
  \caption{The relationship between the distances of densities and distances of wavefunctions for a family of two-electron random-potential systems is seen to be quasi-linear with a gradient of $\sim$1.06 (black dashed line).}
\label{fig:2e_int}
\end{figure}
We generate a family of 10 random potentials with two interacting electrons, using Eq.~\ref{eq:rand_pot} with $\Lambda = 0.1$ and $L = 15$, and then compare every system to every other system in the family using Eqs.~\ref{eq:wf_met} and \ref{eq:den_met}. We obtain the results shown in Fig.~\ref{fig:2e_int} using the iDEA code, which uses the reduced Coulomb interaction: $1/(|x-x'|+1)$.

These two-electron ground-state results display a striking linearity between the wavefunction distances and the density distances, even though the families of systems vary by several parameters (unlike previous studies where families of systems only varied by one parameter, such as the frequency in the Hooke's atom \cite{DAmico}). 

The ratio of this relationship, for these interacting two electron random systems, is $\sim$1.06, consistent with the one of $\sim$1 found in Ref.~\cite{DAmico}, for the isoelectronic Helium series, Hooke's atom and the two electron Hubbard model.  The relationship between the ratio of the distances and the number of electrons, seen in ~\cite{DAmico}, is also observed for unrelated random systems, supporting that this may be a general property for comparisons of any systems.  This can be used to gain insight into the one-to-one mapping between densities and wavefunctions, in turn improving understanding of the Hohenberg-Kohn theorem for DFT.

\subsection{Non-interacting systems}
\label{sec:non-int}
\begin{figure}
\includegraphics[width=0.48\textwidth]{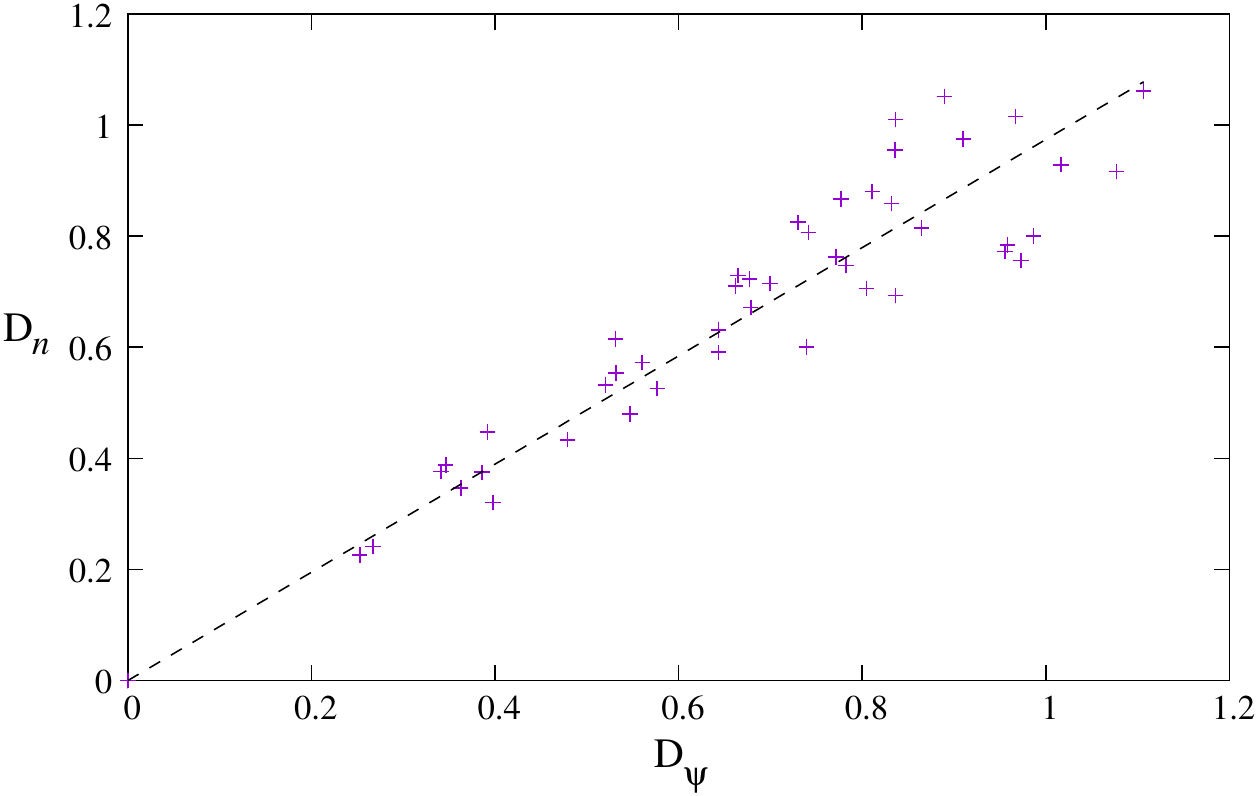}
  \caption{For a family of two electron non-interacting random systems, the relationship between the density distances and wavefunction distances is also seen to be quasi-linear with a gradient of $\sim$0.97 (black dashed line).}
\label{fig:2e_non_int}
\end{figure}
Considering the same family of systems as used in Sec.~\ref{sec:int} but with non-interacting electrons yields the results shown in Fig.~\ref{fig:2e_non_int}.  We see the ratio of distances to be $\sim$0.97, which is very similar to the value seen for the interacting two-electron random-potential systems, and to the value found in \cite{DAmico} for families obtained by varying a parameter in the Hamiltonians.  The striking linearity for both interacting and non-interacting systems opens up many opportunities for these metrics to be used to investigate relationships for a wide range of many-body systems.

The quasi-linear relationship for the ratio of distances remaining in non-interacting systems  enables applications of metrics for Kohn-Sham systems within DFT. Some work in this direction has been done in Ref.~\cite{Sharp2016} and using metrics for Kohn-Sham systems is currently being investigated.

We note that, while the individual points change between the interacting and non-interacting graphs, the overall trend is similar, therefore using this method one can gain information on the effect of interactions for individual systems, as well as on the general trend of families of systems.

\section{Conclusion}
\label{sec:conc}
We have seen how a Fourier series combined with a tailored confining potential, provides a wide range of random potentials with the ability to explore many different quantum systems displaying diverse physics.

From these random potentials, we use metrics to gain insight into the relationship between the  density and wavefunction mapping for time-dependent single electron systems.  This mapping is also at the core of time-dependent DFT \cite{Runge,Marques2006}, where metrics have the potential to play a role at furthering understanding.  We see that the wavefunction distance is impacted greater than the density distance by highly out-of-equilibrium dynamics, and that the quantum systems explored are non-ergodic in metric space.

Extending the use of these random potentials, we have shown that families of two-electron random systems in their ground state still have a quasi-linear relationship between wavefunction distance and density distance, proportional to the number of electrons (the normalisation of the wavefunction has an impact on this relationship, so care needs to be taken when comparing across research).  This is seen to be true in both interacting and non-interacting cases, leading to the possibility of using metrics to investigate Kohn-Sham systems for DFT.

We have seen that Hohenberg-Kohn and Runge-Gross theorems can be described by the metrics for the one and two particle random systems.  This has already highlighted the intricate relationship between wavefunctions and densities in a visual way.  As the number of electrons increases, this relationship becomes more complicated.  Therefore metrics have the potential to provide further insight into, and visually represent, this relationship with the Hohenberg-Kohn theorems for ground states, the Runge-Gross theorems for the time-dependent systems, the link between the ground states and time-dependent systems, and also looking at approximations used for Kohn-Sham systems.

\begin{acknowledgements}
We acknowledge helpful discussions with P. Sharp, J. Wetherell and M. Hodgson, and advice on the iDEA code from J. Wetherell. AHS acknowledges support from EPSRC; IDA acknowledges support from the Conselho Nacional de Desenvolvimento Cientfico e Tecnologico (CNPq, Grant: PVE Processo: 401414/2014-0) and from the Royal Society (Grant no. NA140436).
\end{acknowledgements}

\appendix
\section{System parameters}
\label{sec:append}
Table~\ref{tab:sys_param} gives details of the parameters of Eq.~\ref{eq:rand_pot} for the systems used in Figs. \ref{fig:1e_td}, \ref{fig:2e_int} and \ref{fig:2e_non_int}.  Half the system size, $L$, is 15 a.u., and so $a$ to $f$ represent number drawn from a uniform random distribution from -0.5 to 0.5, where $\Lambda = 0.1$ has been incorporated into the random number distribution for this table.  

The single electron random-potential systems span a range of characteristics which, in the presence of the applied field, range from ballistic motion (including reflections from the system edge) within a broad well, to field-induced tunnelling through a barrier.

\begin{table}[h]
\centering
\label{tab:sys_param}
\begin{tabular}{ |c|c|c|c|c|c|c| }
\hline
 System & $a$ & $b$ & $c$ & $d$ & $e$ & $f$ \\ 
 \hline
 1 e$^-$: 1 & 0.0482 & 0.2373 & 0.2063 & 0.0746 & -0.0322 & -0.2120 \\  
 1 e$^-$: 2 & -0.1042 & 0.0357 & 0.2070 & -0.0966 & 0.2162 & -0.3012 \\  
 1 e$^-$: 3 & -0.2582 & -0.1735 & -0.2054 & -0.1113 & -0.1065 & -0.1126 \\  
 1 e$^-$: 4 & 0.2096 & -0.2042 & -0.1824 & -0.2468 & 0.2638 & 0.1470 \\
 1 e$^-$: 5 & -0.0580 & 0.2908 & -0.2240 & -0.0109 & -0.1114 & -0.1222 \\  
 1 e$^-$: 6 & -0.1050 & -0.1335 & 0.1891 & -0.2625 & -0.2937 & -0.0881 \\  
 1 e$^-$: 7 & -0.1715 & -0.1296 & 0.1299 & -0.2408 & 0.0775 & 0.0179 \\  
 1 e$^-$: 8 & 0.1325 & 0.1391 & 0.2201 & 0.0444 & -0.2139 & 0.1223 \\  
 1 e$^-$: 9 & -0.3019 & 0.0697 & 0.2323 & -0.1791 & 0.2081 & 0.2544 \\  
 1 e$^-$: 10 & -0.2035 & -0.1445 & -0.0631 & 0.0946 & 0.0575 & 0.0542 \\  
 \hline
 2 e$^-$: 1 & 0.0764 & 0.2254 & -0.3165 & -0.1920 & -0.2219 & -0.1148 \\
 2 e$^-$: 2 & 0.0892 & -0.1122 & -0.0649 & -0.2788 & 0.0669 & -0.0780 \\
 2 e$^-$: 3 & -0.1353 & 0.3229 & 0.0182 & 0.2804 & 0.0215 & -0.1230 \\
 2 e$^-$: 4 & -0.1074 & 0.0382 & 0.2487 & -0.1848 & 0.3219 & 0.1215 \\
 2 e$^-$: 5 & 0.1834 & -0.1240 & -0.0136 & -0.0194 & -0.2544 & 0.1890\\
 2 e$^-$: 6 & -0.3318 & 0.3270 & 0.2486 & 0.1503 & 0.2267 & -0.0320 \\
 2 e$^-$: 7 & 0.2972 & 0.1459 & -0.2182 & 0.2943 & -0.2651 & 0.0757 \\
 2 e$^-$: 8 & 0.0433 & 0.2632 & -0.0133 & -0.2655 & -0.1991 & 0.2223 \\
 2 e$^-$: 9 & 0.1315 & -0.0282 & 0.1808 & -0.3280 & 0.1828 & -0.0207 \\
 2 e$^-$: 10 & 0.2003 & -0.1705 & 0.3226 & -0.3230 & 0.0733 & 0.3266 \\
 \hline
\end{tabular}
\caption{Table of the parameters for Eq.~\ref{eq:rand_pot} for the single electron time dependent systems, and the two electron ground state systems.}
\end{table}

\bibliographystyle{spphys}
\bibliography{ref}

\begin{thebibliography}{10}
\providecommand{\url}[1]{{#1}}
\providecommand{\urlprefix}{URL }
\expandafter\ifx\csname urlstyle\endcsname\relax
  \providecommand{\doi}[1]{DOI \discretionary{}{}{}#1}\else
  \providecommand{\doi}{DOI \discretionary{}{}{}\begingroup
  \urlstyle{rm}\Url}\fi

\bibitem{Parr1994}
R.~Parr, W.~Lang, \emph{Density-Functional Theory of Atoms and Molecules}
  (Oxford University Press, Oxford, 1994)

\bibitem{Burke2012}
K.~Burke, The Journal of Chemical Physics \textbf{136}, 150901 (2012)

\bibitem{Hohenberg}
P.~Hohenberg, W.~Kohn, Physical Review \textbf{136 (3B)}, B864 (1964)

\bibitem{DAmico}
I.~D'Amico, J.P. Coe, V.V. Fran\c{c}a, K.~Capelle, Phys. Rev. Lett.
  \textbf{106}, 050401 (2011)

\bibitem{S&D}
P.M. Sharp, I.~D’Amico, Phys. Rev. B \textbf{89}, 115137 (2014)

\bibitem{S&D2}
P.M. Sharp, I.~D’Amico, Phys. Rev. A \textbf{92}, 032509 (2015)

\bibitem{Skelt2017}
A.H. Skelt, R.W. Godby, I.~D'Amico, Phys. Rev. A \textbf{98}, 012104 (2018)

\bibitem{Kohn1965}
W.~Kohn, L.J. Sham, Phys. Rev. \textbf{140}, A1133 (1965)

\bibitem{Herrera2017}
M.~Herrera, R.M. Serra, I.~D'Amico, Scientific Reports \textbf{7}, 4655 (2017)

\bibitem{Hodgson}
M.J.P. Hodgson, J.D. Ramsden, J.B.J. Chapman, P.~Lillystone, R.W. Godby, Phys.
  Rev. B \textbf{88}, 241102(R) (2013)

\bibitem{Note1}
We denote atomic units using a.u., where ${\mathchar '26\mkern -9muh}= m = e =
  4 \pi \varepsilon _0 = 1 $.

\bibitem{Note2}
The Runge-Gross theorem does not prohibit two systems, propagated by (possibly
  different) time-dependent potentials from different initial many-body states
  from having the same density at some later instant. Such a case would provide
  a trajectory that touches the very bottom of the lower triangle in Fig.~\ref
  {fig:1e_td}: $D_n=0$ while $D_{\psi } \not =0$. This provides further support
  for time-evolution in the metric space of Fig.~\ref {fig:1e_td}. being
  predominantly concentrated away from the upper triangle.

\bibitem{Sharp2016}
P.M. Sharp, I.~D'Amico, Phys. Rev. A \textbf{94}, 062509 (2016)

\bibitem{Runge}
E.~Runge, E.K.U. Gross, Phys. Rev. Lett. \textbf{52}, 997 (1984)

\bibitem{Marques2006}
M.~Marques, C.~Ullrich, F.~Nogueira, A.~Rubio, K.~Burke, E.~Gross,
  \emph{Time-dependent density functional theory} (Springer, New York, 2006)

\end{thebibliography}

 
 
 
 
 
 


 

 
 
 


\end{document}